\documentclass[journal]{IEEEtran}

\usepackage{amsmath,amssymb,amsthm}
\usepackage{graphicx}
\graphicspath{{.}}

\usepackage{iftex}
\ifPDFTeX\else
\usepackage{fontspec}
\newcommand{\TGy}{texgyre}
\defaultfontfeatures{Ligatures=TeX}
\edef\PASSrmdefault{\rmdefault}
\DeclareFontFamilySubstitution{TU}{ptm}{\PASSrmdefault}
\fi
\usepackage{algorithm}
\usepackage{algpseudocode}
\usepackage{cite}
\usepackage{xcolor}
\usepackage{placeins}   

\newtheorem{theorem}{Theorem}
\newtheorem{proposition}{Proposition}
\newtheorem{lemma}{Lemma}
\newtheorem{corollary}{Corollary}
\theoremstyle{remark}
\newtheorem{remark}{Remark}

\newcommand{\cM}{\mathcal{M}}
\newcommand{\cN}{\mathcal{N}}
\newcommand{\Ts}{T_{\mathrm{s}}}
\newcommand{\va}{v_{\mathrm{a}}}
\DeclareMathOperator*{\argmax}{arg\,max}

\begin{document}
	
	\title{Actuation-Aware Pinching-Antenna Placement\\ Under Blockage and User Mobility}
	
	\author{Javad Bagherivand}

	\markboth{IEEE Wireless Communications Letters}%
	{Bagherivand: Actuation-Aware PASS Placement}
	
	\maketitle
	
\begin{abstract}
	Existing placement studies for pinching-antenna systems (PASS) usually
	assume that antenna movement is instantaneous and free. In practice,
	actuators move at finite speeds and must pause radiation while
	repositioning. This makes performance depend on consecutive antenna
	positions, turning placement into an edge-weighted longest-path problem on
	a temporal trellis. We also show that an obstacle blocks a wider range of
	antenna positions as it approaches the user, causing sudden performance
	drops that make multi-step lookahead essential, and that once obstacle
	height is included every blockage event shadows at least a fixed minimum
	length of waveguide. We give an exact offline solution with complexity
	$\mathcal{O}(TMS)$ and bound the per-slot loss of a $W$-step
	receding-horizon policy by $\mathcal{O}(1/W)$. Measurements show that the
	required lookahead window is set by shadow width rather than by waveguide
	length. Across $20$ random seeds, the exact policy delivers $17.7\%$ higher
	max-min throughput than per-slot optimization under blockage, and matches
	it when no blockers are present.
\end{abstract}
	
	
	\vspace{-4pt}
\section{Introduction}
\IEEEPARstart{P}{inching-antenna} systems (PASS) place small dielectric
elements at chosen points along a dielectric waveguide. This lets the
network create line-of-sight links wherever it wants, instead of fixing
antenna positions at deployment time \cite{fukuda2022,ding2025flexible}.
Since the channel depends directly on the antenna's coordinate, placement
optimization becomes the main design problem. A large body of work studies
this jointly with beamforming and multiple access
\cite{xu2025power,tegos2025rate,xie2025lowcomplexity,zhao2026wdma}. PASS
differs from reconfigurable intelligent surfaces in one key way: RIS
reshapes the channel using electronic phase weights at a fixed location
\cite{wu2020towards}, while PASS moves the radiating element itself
\cite{li2024movementdelay}. That difference drives everything that follows.
A moving aperture couples neighboring slots and widens the shadow a blocker
casts. A moving phase profile does neither.

The placement literature relies on two assumptions: the antenna can reach
any point instantly, and the design is recomputed for every channel
snapshot. Mobility-aware approaches, mostly based on deep reinforcement
learning \cite{drl2026mobility,sac2026umi}, drop the second assumption but
keep the first --- they treat reconfiguration time as negligible compared to
a one-second control interval. The cost of that is already documented.
Mechanical delay limits how freely the element can move
\cite{constrained2026}; how often you reconfigure trades off against
transmission overhead \cite{dsp2026trajectory}; and movement delay eats
into usable transmission time in movable-antenna systems
\cite{li2024movementdelay}. Each time it shows up only as a caveat. No
existing formulation treats the actuator constraint as a design variable.

This gap matters. A finite actuator speed couples consecutive placements
together, which breaks the per-snapshot decomposition that every existing
algorithm depends on. If radiation is interrupted while the element
repositions, the slot utility becomes a function of the \emph{transition}
between positions, not just the position itself. This turns a
node-weighted selection problem into an edge-weighted one. Trellis search
over discrete activation states has been applied to PASS before
\cite{viterbi2025}, but only on node weights --- the exact structure that a
finite actuator invalidates.

We also make a second observation, specific to PASS, that appears to be
new. Because the aperture slides, an obstacle blocks an interval of
antenna positions whose width grows as $1/(1-y_b/y_k)$
(Lemma~\ref{lem:shadow}), so an obstacle tens of centimeters wide can
shadow meters of waveguide once it nears the \emph{user}. This sharpens
the known result that blockage in PASS can help or hurt depending on
geometry \cite{dingpoor2025blockage}. Obstacle height removes the narrow
shadows, so every blockage event shadows at least $w_b\,d/H_b$. These
wide notches sweep through the utility landscape, so anticipating them
pays; without blockers, per-slot optimization is already optimal
(Sec.~\ref{sec:sim}).

\textbf{Contributions.} (i) We formulate an actuation-aware horizon
problem with a slew-rate constraint and a transition-dependent radiation
duty factor, and show it is an edge-weighted trellis problem
(Sec.~\ref{sec:model}). (ii) We derive the shadow-width law
(Lemma~\ref{lem:shadow}), its height-constrained sharpening (no blockage
event is ever benign), and the transit law \eqref{eq:tauB}, where this
width amplification cancels out. (iii) We prove that myopic optimization --- the implicit policy in the snapshot literature --- is
either infeasible or strictly suboptimal (Prop.~\ref{prop:myopic}), and we
give an exact $\mathcal{O}(TMS)$ dynamic program (Thm.~\ref{thm:dp}).
(iv) We derive a $1/W$ prediction-window bound (Thm.~\ref{thm:rh},
Cor.~\ref{cor:rule}) and show through measurements that shadow width,
not waveguide length, is what actually sets this window in
practice.

	\vspace{-4pt}
\section{System Model}\label{sec:model}
A ceiling-mounted waveguide carries one movable radiating element that
serves $K$ ground users, while moving obstacles block some of the links.
Below we describe the geometry and channel, the limits on how the element
moves, and the throughput each slot delivers.
\subsection{Geometry, Channel and Blockage}
A waveguide of length $L$ lies along the $x$-axis at ceiling height $d$, fed
at $x=0$. $K$ single-antenna users move around in the covered region; user
$k$ is at $(x_k(t),y_k(t),0)$ in slot $t$. Time is slotted with duration
$\Ts$ over a horizon $T$. We follow the discrete-activation model motivated
by practical actuation hardware \cite{xu2025power,uplinkact2026}: the
waveguide has $M$ preconfigured positions $\chi_m=(m-1)\Delta$,
$\Delta=L/(M-1)$, and only one antenna is active at a time, indexed
$m_t\in\cM\triangleq\{1,\dots,M\}$.

\begin{figure}[t]\centering
	\includegraphics[width=3.4in]{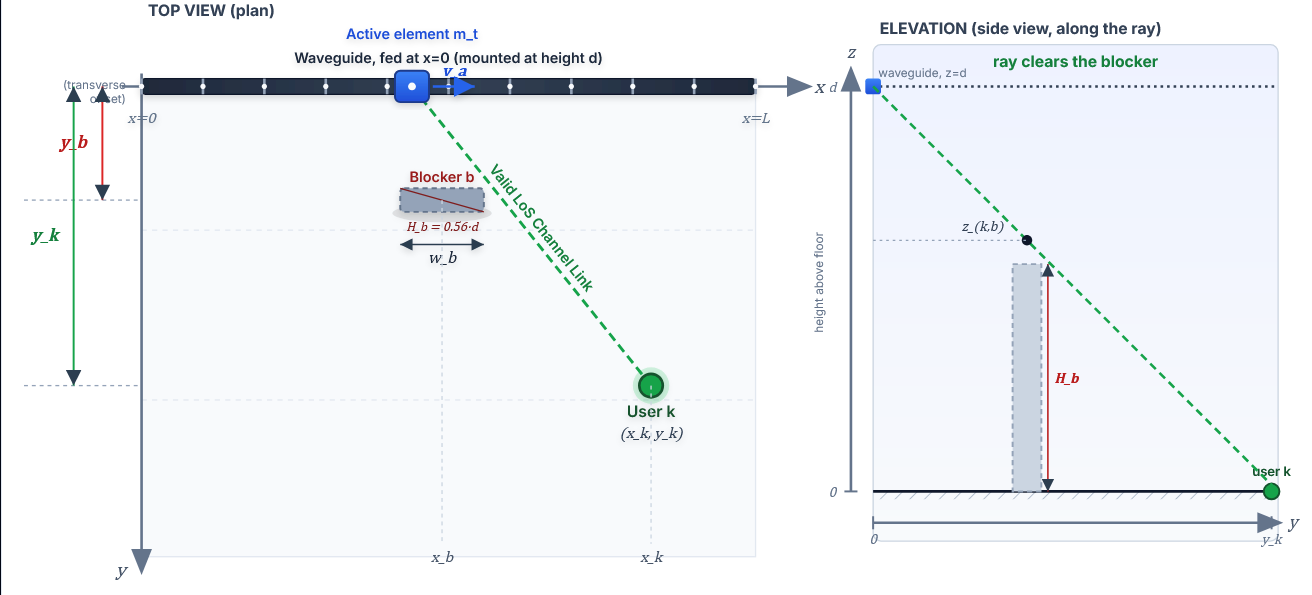}
	\caption{Geometry of the actuation-aware PASS model. \emph{Left:} plan view,
	showing waveguide (fed at $x=0$, height $d$), active element $m_t$
	(max speed $\va$), user $k$, and blocker $b$'s footprint at offset $y_b$.
	The active position avoids the shadow interval of
	Lemma~\ref{lem:shadow}, so the ray (dashed) reaches the user.
	\emph{Right:} elevation view --- the ray clears the blocker since its
	height $z_{k,b}=d(1-y_b/y_k)$ exceeds $H_b$. Occlusion requires both the
	plan-view test and the height test \eqref{eq:height} to
	fail.}\label{fig:sys}
\end{figure}

$N_B$ blockers of width $w_b$ and height $H_b$ sit at $(x_b(t),y_b(t))$.
Fig.~\ref{fig:sys} shows the geometry in the two views the model needs.
Consider the ray from the antenna at $(\chi,0,d)$ to user $k$ at
$(x_k,y_k,0)$: it reaches the blocker plane $y=y_b$ at $s=y_b/y_k$. Two
independent conditions must both hold for the link to be blocked, and each
one shows up in a different view. In plan view (Fig.~\ref{fig:sys}, left)
the obstacle must sit \emph{between} the waveguide and the user, so
$0<y_b<y_k$, and the ray must actually cross its footprint. The ray meets
$y_b$ at abscissa
\begin{equation}\label{eq:ray}
	\xi_{k,b}(\chi)=\chi+\tfrac{y_b}{y_k}\,(x_k-\chi),
\end{equation}
so this first test reads $|\xi_{k,b}(\chi)-x_b|<w_b/2$. In elevation
(Fig.~\ref{fig:sys}, right) the obstacle must also be tall enough to catch
the ray at the point where it passes. The ray height there is
\begin{equation}\label{eq:height}
	z_{k,b}=d\,(1-y_b/y_k)\ \le\ H_b.
\end{equation}
If either test fails, the link stays line-of-sight, which is the case shown
in the figure.

\begin{lemma}[Shadow-width law]\label{lem:shadow}
	For $0<y_b<y_k$ satisfying \eqref{eq:height}, the antenna positions
	blocked for user $k$ by blocker $b$ form an interval centered at
	$\bar\chi_{k,b}=(x_b-\tfrac{y_b}{y_k}x_k)/(1-\tfrac{y_b}{y_k})$, with width
	\begin{equation}\label{eq:shadow}
		w^{\mathrm{sh}}_{k,b}=\frac{w_b}{1-y_b/y_k}\ \ge\ \frac{w_b\,d}{H_b},
	\end{equation}
	clipped to the waveguide extent $[0,L]$.
\end{lemma}

\textit{Proof:} The interval follows by inverting
$|\xi_{k,b}(\chi)-x_b|<w_b/2$, since $\xi_{k,b}(\cdot)$ is affine in
$\chi$ with slope $1-y_b/y_k\in(0,1)$. The lower bound comes from
$1-y_b/y_k\le H_b/d$, which \eqref{eq:height} forces.\hfill\IEEEQED

As $y_b\to y_k$ the shadow blows up: a $0.6$\,m obstacle sitting
$0.2$\,m from a user $4$\,m out ($y_b/y_k=0.95$) shadows $12$\,m of
antenna positions, and the sweep in Sec.~\ref{sec:sim} averages
$4.6$\,m. This flips the usual intuition that blockers near the user
are the mild case --- here they are the worst case. The height test
\eqref{eq:height} then removes exactly the \emph{narrow} shadows. A
blocker close to the waveguide, which \eqref{eq:shadow} would otherwise
credit with a shadow of only $\approx w_b$, does not block anything at
all: the ray simply passes over it. Every event that survives shadows
at least $w_b\,d/H_b$, i.e., $1.06$\,m for $w_b=0.6$\,m, $d=3$\,m,
$H_b=1.7$\,m; a user at $y_k=4$\,m cannot even be blocked by obstacles
nearer than $y_k(1-H_b/d)=1.73$\,m. In short, there are no benign
blockage events in PASS. Raising the waveguide height widens the
minimum shadow while narrowing the band of positions that block at
all.

These shadows do not sit still, and how fast they move matters as much as
how wide they are. Let $\rho\triangleq y_b/y_k$ and hold it fixed, as the
mobility model in Sec.~\ref{sec:sim} does. The shadow center from
Lemma~\ref{lem:shadow} then sweeps at rate
$\dot{\bar\chi}_{k,b}=(\dot x_b-\rho\,\dot x_k)/(1-\rho)$, so a shadow
clears its own width in time
\begin{equation}\label{eq:tauB}
	\tau_B=\frac{w^{\mathrm{sh}}_{k,b}}{|\dot{\bar\chi}_{k,b}|}
	=\frac{w_b}{|\dot x_b-\rho\,\dot x_k|}\,;
\end{equation}
note that the amplification factor \emph{cancels out}. (If $\rho$
drifts, it adds $\dot\rho\,(x_b-x_k)/(1-\rho)^2$ to
$\dot{\bar\chi}_{k,b}$.) A blocker near the user casts a shadow ten
times wider, but also sweeps it ten times faster. As a result, the dwell time at
any one antenna position ends up fixed by the obstacle's physical
width and the relative in-plane velocity --- nothing else. This is why
the notches are wide but the prediction window in Sec.~\ref{sec:rh}
can still be short.

Let $r_k(m,t)=\sqrt{(\chi_m-x_k(t))^2+y_k(t)^2+d^2}$, and let the blockage
indicator be $\varsigma_k(m,t)\in\{1,\varsigma_{\mathrm{b}}\}$ with
$\varsigma_{\mathrm{b}}\ll1$. The received power gain is then
\begin{equation}\label{eq:channel}
	|h_k(m,t)|^2=\frac{\eta\,e^{-2\alpha\chi_m}\,\varsigma_k(m,t)}{r_k^2(m,t)},
\end{equation}
where $\eta=c^2/(4\pi f_c)^2$ and $\alpha$ is the in-waveguide attenuation
coefficient \cite{dualfed2026}, given in Np/m --- a value quoted in dB/m
must first be divided by $20/\ln 10$ before it enters \eqref{eq:channel}.
The indicator is binary, so overlapping shadows attenuate a position
only once, which understates the benefit of avoiding them. With one active
element the guided-phase term drops out of \eqref{eq:channel}; the
landscape is multimodal because of blockage notches and spread-out users.

\subsection{Actuation}
Repositioning is mechanical, with a maximum speed $\va$ and a settling
time $\tau_0$. Within one slot the element can move at most $\va\Ts$,
which gives
\begin{equation}\label{eq:S}
	S\triangleq\lfloor \va\Ts/\Delta\rfloor,\quad
	m_t\in\cN_S(m_{t-1})\triangleq\{m\!:\!|m-m_{t-1}|\le S\}.
\end{equation}
While the element is translating and settling, its coupling to the
guided mode is transient and the link cannot be used, so the usable
fraction of slot $t$ is
\begin{equation}\label{eq:beta}
	\beta(m',m)=\Big(1-\tfrac{|m-m'|\Delta/\va+\tau_0\mathbf{1}\{m\neq m'\}}{\Ts}\Big)^{\!+}.
\end{equation}
Setting $\tau_0=0$ and letting $\va\to\infty$ gives $\beta\equiv1$ and
$S\ge M-1$. This recovers the idealized model used in
\cite{tegos2025rate,xie2025lowcomplexity,drl2026mobility}, which we keep
below as an upper bound labeled ``Ideal.''

\subsection{Utility}
Only one element radiates at a time, so all $K$ users are served from the
same coordinate $\chi_{m_t}$ and share the slot through equal-time TDMA.
Each user gets a fraction $1/K$ of the usable time $\beta\Ts$, at whatever
rate its own link supports from that single position. This means the slot
is only as good as its \emph{weakest} link. The max-min throughput
delivered in slot $t$, under this equal split and the transition
$m_{t-1}\!\to\!m_t$, is
\begin{equation}\label{eq:edge}
	g_t(m_{t-1},m_t)=\frac{\beta(m_{t-1},m_t)}{K}
	\min_{k}\log_2\!\Big(1+\tfrac{P|h_k(m_t,t)|^2}{\sigma^2}\Big).
\end{equation}
This letter rests on one key fact: $g_t$ depends on the ordered
\emph{pair} $(m_{t-1},m_t)$, not on $m_t$ alone, so it is an edge weight
on a trellis, whereas snapshot formulations produce only node weights.
Thms.~\ref{thm:dp} and \ref{thm:rh} need only a bounded slot reward of
consecutive placements, so sum-rate or weighted-fair objectives work as
well. The objective changes the \emph{size} of the gain, not its
existence; max-min stresses a shared aperture hardest, since one shadowed
user caps the whole slot.

	\vspace{-4pt}
\section{Problem and the Failure of Myopic Placement}
\begin{equation}\label{eq:P1}
	\textbf{(P1)}\ \max_{\{m_t\}\in\cM^{T}}\ \sum_{t=1}^{T}g_t(m_{t-1},m_t)
	\quad\text{s.t.}\quad m_t\in\cN_S(m_{t-1}),
\end{equation}
with $m_0$ given. (P1) searches over $M^T$ sequences. Let
$u_t(m)\triangleq g_t(m,m)$, and let $m^{\mathrm{my}}_t=\argmax_m u_t(m)$
be the per-snapshot optimizer that the existing literature implicitly
targets.

\begin{proposition}\label{prop:myopic}
	(i) If $|m^{\mathrm{my}}_t-m_{t-1}|>S$, the myopic placement violates
	\eqref{eq:S} and cannot be realized. (ii) Even when $S\ge M-1$, the
	myopic policy can still be strictly suboptimal, with a gap that grows
	linearly in $T$.
\end{proposition}
\textit{Proof:} (i) is immediate. For (ii), take $M=3$, $S=2$, with
$u_t(2)=a$ for all $t$, $u_t(1)=a+\delta$ for odd $t$, and $u_t(3)=a+\delta$
for even $t$ (the remaining entry stays $a$). Write $\bar\beta<1$ for the
value of $\beta$ at $|m-m'|=2$. The myopic policy alternates
$1\!\to\!3\!\to\!1$, collecting $\bar\beta(a+\delta)$ every slot after the
first. Staying at $2$ instead collects $a$. For
$\delta<a(1-\bar\beta)/\bar\beta$, the stationary policy wins by
$(T-1)[a-\bar\beta(a+\delta)]$, which grows linearly in $T$.\hfill\IEEEQED

This bad case needs the utility peak to jump between distant positions.
Blockage produces exactly that kind of jump; smooth mobility does not
(Sec.~\ref{sec:model}). This is why Sec.~\ref{sec:sim} shows a
substantial gap with blockers present, and none without them.

\vspace{-4pt}
\section{Exact Offline Solution}\label{sec:offline}
\begin{theorem}\label{thm:dp}
	(P1) is a longest-path problem on a directed acyclic graph (DAG) with
	$TM$ nodes, where node $(t,m)$ connects to $(t{+}1,m')$ for
	$m'\in\cN_S(m)$ with edge weight $g_{t+1}(m,m')$. The global optimum
	follows from
	\begin{equation}\label{eq:bellman}
		V_{t-1}(m')=\max_{m\in\cN_S(m')}\{g_t(m',m)+V_t(m)\},\quad V_T(\cdot)=0,
	\end{equation}
	giving $J^\star=V_0(m_0)$, in $\mathcal{O}(TM(2S{+}1))$ time and
	$\mathcal{O}(TM)$ memory.
\end{theorem}

\textit{Proof:} Both the constraint and the objective depend only on
consecutive pairs, so $m_{t-1}$ is a sufficient state and
\eqref{eq:bellman} is exactly the Bellman recursion on a finite DAG. Each
of the $TM$ nodes has at most $2S{+}1$ outgoing edges.\hfill\IEEEQED

Brute-force search costs $M^T$, so Thm.~\ref{thm:dp} takes us from
exponential to linear cost with no loss of optimality --- unlike the
successive-convex and heuristic placement methods in
\cite{tegos2025rate,xie2025lowcomplexity}, which only guarantee a
stationary point. Actuation hardware only exposes finitely many
positions anyway \cite{uplinkact2026}, so the grid problem \emph{is} the
real physical problem.

\vspace{-4pt}
\section{Causal Operation and the Prediction Window}\label{sec:rh}
\begin{figure}[t]\centering
	\includegraphics[width=3.3in]{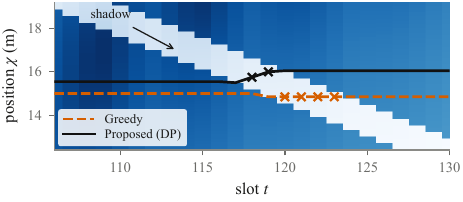}
	\caption{One measured shadow crossing ($K=1$, $N_B=4$, nominal actuator,
		$0.25$\,m per slot). Light cells are shadowed positions; crosses mark
		slots spent in shadow.}\label{fig:mech}
\end{figure}

Thm.~\ref{thm:dp} needs the whole trajectory in advance. In practice the
scheduler only predicts $W$ slots ahead (Alg.~\ref{alg:rh}).
Fig.~\ref{fig:mech} shows why looking ahead matters. The shadow sweeps
about as fast as the element can move, so no policy can outrun it. Greedy
steps the same way and is blocked for four slots; the offline optimum of
Thm.~\ref{thm:dp} crosses behind it and loses two. Alg.~\ref{alg:rh} with
$W\ge3$ loses two or three.

We assume
(A1) $S\ge1$; (A2) $g_t\in[g_{\min},g_{\max}]$, $G\triangleq
g_{\max}-g_{\min}$; (A3) predictions are exact within the window. Let
\begin{equation}\label{eq:n0}
	n_{\mathrm{geo}}\triangleq\big\lceil (M-1)/S\big\rceil
\end{equation}
be the number of slots after which every grid position becomes reachable
from every other, since the $n$-step reachable set spans $\cM$ once
$nS\ge M-1$. This equals the traversal time $\lceil L/(\va\Ts)\rceil$
when $\va\Ts$ is a whole multiple of $\Delta$, and exceeds it otherwise,
since the floor in \eqref{eq:S} loses a little.

\begin{lemma}[Value-span bound]\label{lem:span}
	Let $V^{(h)}_t(m)$ be the optimal reward-to-go over $h$ remaining
	slots from state $m$. Then $|V^{(h)}_t(m)-V^{(h)}_t(m')|\le
	n_{\mathrm{geo}}G$ for all $m,m'$.
\end{lemma}
\textit{Proof:} For $h\le n_{\mathrm{geo}}$ this is immediate: both values
are sums of $h$ rewards drawn from a range of width $G$, so they differ by
at most $hG\le n_{\mathrm{geo}}G$, and no reachability argument is needed.
For $h>n_{\mathrm{geo}}$, the $n$-step reachable set
$[m-nS,m+nS]\cap\cM$ equals $\cM$ at $n=n_{\mathrm{geo}}$ by
\eqref{eq:n0}. Given an optimal path $p$ from $m'$, the path from $m$ that
reaches $p_{n_{\mathrm{geo}}}$ and matches $p$ afterward is feasible, and
it differs from $p$ in at most $n_{\mathrm{geo}}$ slot rewards, each by at
most $G$. This gives $V^{(h)}_t(m)\ge V^{(h)}_t(m')-n_{\mathrm{geo}}G$,
and swapping $m$ with $m'$ gives the other
direction.\hfill\IEEEQED

\begin{theorem}[Prediction-window bound]\label{thm:rh}
	Let $\pi_W$ solve a $W$-stage instance of \eqref{eq:bellman} from the
	current state, commit its $W$ actions, and re-solve. Then
	\begin{equation}\label{eq:thm2}
		J^\star-J^{\pi_W}\ \le\ n_{\mathrm{geo}}\,G\Big(\tfrac{T}{W}+1\Big),
	\end{equation}
	i.e., the per-slot gap is at most $n_{\mathrm{geo}}G(1/W+1/T)$.
\end{theorem}
\textit{Proof:} Split the horizon into $B=\lceil T/W\rceil\le T/W+1$
blocks, each of length $W$ except possibly the last, and let
$\{m^\star_t\}$ be the offline optimum. Write $W_b$ for the length of
block $b$ and $J_b$ for the reward it collects. Since $\pi_W$ is optimal
over that block starting from its entry state $m_b$, we have
$J^{\pi_W}_b=V^{(W_b)}_b(m_b)$, while the offline optimum restricted to
the block satisfies $J^\star_b\le V^{(W_b)}_b(m^\star_b)$.
Lemma~\ref{lem:span} bounds the difference by $n_{\mathrm{geo}}G$ per
block, for any $W_b$. Summing over the $B$ blocks gives
\eqref{eq:thm2}.\hfill\IEEEQED

\begin{corollary}\label{cor:rule}
	A per-slot gap of at most $\varepsilon$ is guaranteed by
	$W\ge 2n_{\mathrm{geo}}G/\varepsilon$, as long as
	$T\ge 2n_{\mathrm{geo}}G/\varepsilon$.
\end{corollary}

\begin{remark}[What actually sets the window]\label{rem:binds}
	On the nominal grid of Sec.~\ref{sec:sim} ($M=601$, $\Ts=0.5$\,s,
	$\va=0.5$\,m/s, so $S=5$), \eqref{eq:n0} gives $n_{\mathrm{geo}}=120$
	slots. The bound \eqref{eq:thm2} is therefore loose: it charges the policy
	as if it had to cross the whole waveguide, and at $T=250$ its residual term
	$n_{\mathrm{geo}}G/T$ alone is $0.48\,G$. What matters is how long a shadow
	sits on the element's position. By \eqref{eq:tauB} this is
	$\tau_B=w_b/|\dot x_b-\rho\,\dot x_k|$, about $0.9$\,s or
	$n_{\mathrm{blk}}\triangleq\lceil\tau_B/\Ts\rceil\approx2$ slots; the DP in
	Fig.~\ref{fig:mech} spends exactly two slots crossing one. We do not claim
	\eqref{eq:thm2} holds with $n_{\mathrm{blk}}$ in place of
	$n_{\mathrm{geo}}$: that would need $u_t$ and $u_{t+n}$ to decorrelate for
	$n\ge n_{\mathrm{blk}}$, which bounds only the \emph{expected} entry-state
	advantage, not the pathwise one Lemma~\ref{lem:span} needs. Our claim here
	is empirical. Sec.~\ref{sec:sim} measures a $95\%$ knee at
	$W^\star\approx9$ slots, roughly $n_{\mathrm{blk}}$ scaled by the
	$2G/\varepsilon$ factor of Cor.~\ref{cor:rule}, and $W^\star$ tracks $w_b$
	while staying nearly flat in $\va$, blocker speed and user speed. Shadow
	width, not waveguide length, sets the window. The exponential decay in $W$
	known for strongly convex smooth costs \cite{lin2021perturbation} does not
	carry over to a finite, notched landscape.
\end{remark}

\begin{remark}[What is proved versus what is run]\label{rem:commit}
	Thm.~\ref{thm:rh} analyzes a policy that commits all $W$ actions before
	re-solving, because the telescoping argument needs that block structure.
	Alg.~\ref{alg:rh} commits \emph{one} action and re-solves every slot, as a
	real scheduler would. These are different policies, and \eqref{eq:thm2} is
	proved only for the first; since block-committing MPC is not monotone in
	$W$, it cannot bound the second. Measured, the one-step policy is better by
	$1$--$2$ points of the offline optimum at every $W$ (Sec.~\ref{sec:sim}),
	so \eqref{eq:thm2} is the conservative statement. That ordering is an
	empirical finding, not a theorem.
\end{remark}

\begin{remark}[On (A3)]\label{rem:pred}
	Assumption (A3) grants exact user and blocker states over the window, so
	\eqref{eq:thm2} separates the cost of a \emph{finite} horizon from that of
	an \emph{inaccurate} one. A misplaced shadow edge changes the reward by at most $G$ per affected
	slot, so prediction errors add at most $G$ times the expected number of
	mispredicted slots to \eqref{eq:thm2}. We do not quantify that term; the
	results below are the no-prediction-error ceiling for any causal scheme.
\end{remark}

\begin{algorithm}[t]
	\caption{Receding-horizon actuation-aware placement}\label{alg:rh}
	\small
	\begin{algorithmic}[1]
		\State \textbf{Input:} $m_0$, window $W$, grid $\{\chi_m\}$
		\For{$t=1$ \textbf{to} $T$}
		\State Predict user/blocker states over $[t,t{+}W{-}1]$; form $g_{t+i}$
		from \eqref{eq:ray}--\eqref{eq:edge}
		\State $V_W\!\gets\!0$; \textbf{for} $i=W$ \textbf{down to} $1$: apply \eqref{eq:bellman}
		\State $m_t\gets\argmax_{m\in\cN_S(m_{t-1})}\{g_t(m_{t-1},m)+V_1(m)\}$
		\State Actuate to $\chi_{m_t}$; transmit for $\beta(m_{t-1},m_t)\Ts$\,s
		\EndFor
	\end{algorithmic}
\end{algorithm}

\vspace{-4pt}
\section{Numerical Results}\label{sec:sim}

\begin{figure}[!t]\centering
	\includegraphics[width=3.3in]{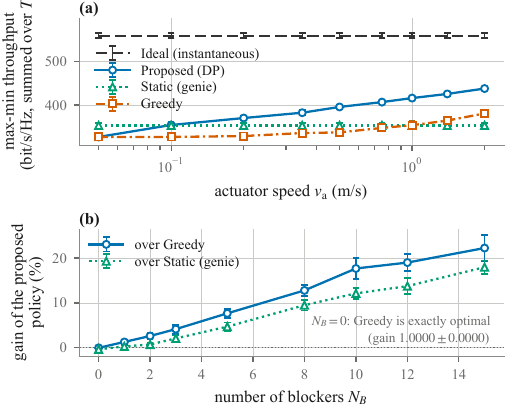}
	\caption{(a) Max-min throughput versus actuator speed: the gap
		between ``Ideal'' and every realizable scheme is the error you
		get from assuming instantaneous repositioning. (b) Gain over
		Greedy and over Static versus blocker density. At $N_B=0$ the
		gain is exactly zero --- without blockage the myopic policy is
		already optimal and the trellis structure does nothing.}\label{fig:va}
\end{figure}

\begin{figure}[!t]\centering
	\includegraphics[width=3.3in]{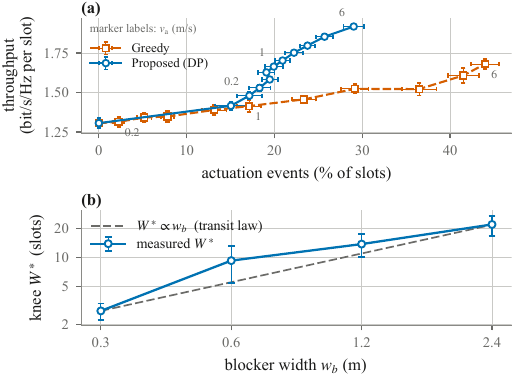}
	\caption{(a) Throughput against actuation cost, one marker per
		actuator speed $\va$ (labels in m/s), with $95\%$ intervals on
		both axes. The proposed policy gains throughput at every speed
		while its movement rate stays nearly flat.
		(b) The $95\%$ knee $W^\star$ versus blocker width, with a
		$\propto w_b$ reference: what sets the window is the width of
		the shadow the element has to clear.}\label{fig:W}
\end{figure}

\textbf{Setup.} We use a $30\!\times\!6$\,m corridor, waveguide at
$d=3$\,m, $f_c=28$\,GHz, $\alpha=0.1$\,dB/m (i.e., $0.011513$\,Np/m in
\eqref{eq:channel}), $\Delta=5$\,cm ($M=601$), $\Ts=0.5$\,s, $T=250$,
$K=4$ users moving at $1$--$3$\,m/s, $N_B=10$ blockers of width
$0.6$\,m and height $1.7$\,m moving at $0.8$--$1.5$\,m/s, blockage loss
$25$\,dB, $P=20$\,dBm, $\sigma^2=-90$\,dBm, $\tau_0=10$\,ms,
$\va=0.5$\,m/s unless we sweep it; $m_0$ is the grid center. Blockers
sit beside the users, with $y_b=\rho\,y_k$ where
$\rho\sim\mathcal{U}(0.55,0.90)$, and pick headings independently of
them; we test both choices below. All results are means over $20$
paired seeds with $95\%$ confidence intervals. Since no actuator speed
or settling time has been published for PASS hardware, we sweep $\va$
over $1.6$ decades rather than commit to a single value, and show that
$\tau_0$ barely matters. Baselines: \emph{Ideal} (free instantaneous
repositioning, the model used in
\cite{tegos2025rate,drl2026mobility}); \emph{Greedy}, the best
reachable transition each slot; \emph{Static}, the best fixed position
with free travel to it, i.e., a genie-aided upper bound on never moving. We
report no learned baseline:
$J^\star$ is exact and upper-bounds every policy,
\cite{sac2026umi} included.

\textbf{Findings.} The idealized model overstates achievable max-min
throughput by $41\%$ ($J^\star/J^{\mathrm{ideal}}=0.707\pm0.009$), so
gains that assume instantaneous repositioning are not
achievable. Against realizable schemes, our policy gains
$17.7\pm2.3\%$ over Greedy and $12.1\pm1.3\%$ over Static
(Fig.~\ref{fig:va}(a)). The shadow geometry of Sec.~\ref{sec:model} explains both
numbers: at $N_B=0$ the gain over Greedy is exactly zero (measured
ratio $1.0000\pm0.0000$), and Static lands within $0.4\%$ of optimal
(Fig.~\ref{fig:va}(b)). Without blockage, the trellis structure does
nothing and per-slot optimization is already optimal, which marks when
the method is not worth its complexity. The gain climbs steadily with blocker density and with $K$
($5.2\%$, $9.2\%$, $17.7\%$, $25.5\%$ for $K=1,2,4,8$). Robustness checks
leave it essentially unchanged: it varies by under one point across user
speeds of $0.5$--$4$\,m/s and settling times $\tau_0\le100$\,ms, survives
at $6.7$--$9.9\%$ when blockers are scattered uniformly rather than placed
beside the users, and shifts by under $0.6\%$ when $\Delta$ is halved. Blockage
therefore drives the gain, and mobility does not.

\begin{table}[t]
	\caption{Cost and actuation wear at the nominal point ($\va=0.5$\,m/s),
		mean over $20$ paired seeds. Throughput is per slot; travel and events are
		per $250$-slot horizon. Runtime is one horizon on one CPU core. Gains
		quoted in the text are means of per-seed ratios, so they differ slightly
		from ratios of these means.}
	\label{tab:cost}
	\centering\footnotesize
	\setlength{\tabcolsep}{3.2pt}
	\begin{tabular}{lccccc}
		\hline
		& Complexity & Thr. & Travel & Events & Runtime\\
		& & {\scriptsize bit/s/Hz} & {\scriptsize m} & {\scriptsize \% slots} & \\
		\hline
		Proposed & $\mathcal{O}(TM(2S{+}1))$ & $1.583$ & $9.3$ & $19.5$ & $16.7$\,ms\\
		Greedy   & $\mathcal{O}(T(2S{+}1))$  & $1.348$ & $2.2$ & $7.8$  & $1.0$\,ms\\
		Static   & $\mathcal{O}(TM)$         & $1.412$ & $0$   & $0$    & ---\\
		Ideal    & $\mathcal{O}(TM)$         & $2.240$ & ---   & ---    & ---\\
		\hline
	\end{tabular}
\end{table}

Actuation cost separates the schemes in a way throughput alone does
not (Table~\ref{tab:cost}). At $\va=0.5$\,m/s our policy repositions
in $19.5\%$ of slots and Greedy in $7.8\%$: moving is expensive, so the
myopic policy mostly stays put and gives up throughput. Static even
beats Greedy here ($1.412$ versus $1.348$), so chasing the best
reachable transition is worse than never moving. Our Greedy is
$\beta$-aware and thus stronger than the node-weight policy
Prop.~\ref{prop:myopic} refutes; the mechanism is the same myopia, but
this measurement is not an instance of the proposition. Computation is cheap too: one horizon takes
$16.7$\,ms ($67\,\mu$s per slot) on a single CPU core, with no GPU.

Fig.~\ref{fig:W}(a) shows both costs at once. The curves start together
at $\va=0.05$\,m/s, where the element cannot cover a grid step per slot
and neither policy moves. From there our policy delivers more throughput
at every speed, while its movement rate stays at $17$--$22\%$ of slots
for $\va=0.2$--$2$\,m/s and Greedy's climbs from $2\%$ to $29\%$. Our
policy moves only when a move pays off over the window; Greedy moves
whenever one looks good in the current slot. As a result, beyond
$\va\approx1.2$\,m/s our policy actuates \emph{less often} than Greedy
and still delivers $14$--$19\%$ more throughput ($0.65\times$ the
actuation events at $\va=2.8$\,m/s).

For the causal policy, Alg.~\ref{alg:rh} reaches $85.1\%$, $91.8\%$, $94.3\%$ and $95.8\%$ of
the offline optimum at $W=1,4,8,16$; the
variant that commits all $W$ actions, which is what
Thm.~\ref{thm:rh} analyzes, runs $1$--$2$ points lower throughout.
The $95\%$ knee sits at $W^\star=9.3\pm3.9$ slots and stays inside its
confidence interval as $\va$ spans $0.2$--$1.0$\,m/s ($8.6$, $9.3$,
$6.9$), which rules out $n_{\mathrm{geo}}$ as the driving factor,
since that would predict $60$--$300$ slots. Shadow width sets it
instead. Fig.~\ref{fig:W}(b) sweeps $w_b$ over $8\times$ and
$W^\star$ tracks it closely: $2.8\to9.3\to13.8\to22.2$ slots, while
blocker speed ($0.25$--$3$\,m/s) and user speed ($0.5$--$4$\,m/s)
each leave $W^\star$ nearly flat. The measured gap decays closer to
$W^{-1/2}$ than to $W^{-1}$ over $W\in[1,16]$, so \eqref{eq:thm2}
holds without being tight.

\FloatBarrier
\vspace{-6pt}
\section{Conclusion}
Treating pinching-antenna repositioning as instantaneous overstates
achievable throughput by $41\%$ and produces policies that cannot be
realized. Once actuator speed and radiation interruption are modelled,
slot utility depends on the transition between positions, which turns
placement into an edge-weighted trellis problem with an exact
linear-time solution. Blockage is what makes anticipation worth its
cost, and once obstacle height is accounted for, no blockage event is
benign. The required prediction window follows shadow width, not waveguide
length.


\vspace{-4pt}
	{\footnotesize
		\begin{thebibliography}{99}
			\setlength{\itemsep}{0pt}\setlength{\parsep}{0pt}\setlength{\topsep}{0pt}
			\bibitem{fukuda2022} A.~Fukuda \emph{et al.}, ``Pinching antenna---using a
			dielectric waveguide as an antenna,'' \emph{NTT DOCOMO Tech. J.}, vol.~23,
			no.~3, pp.~5--12, 2022.
			\bibitem{ding2025flexible} Z.~Ding, R.~Schober, and H.~V. Poor,
			``Flexible-antenna systems: A pinching-antenna perspective,'' \emph{IEEE
				Trans. Commun.}, vol.~73, no.~10, pp.~9236--9253, 2025.
			\bibitem{xu2025power} X.~Xu \emph{et al.}, ``PASS: Power radiation model and
			optimal beamforming design,'' \emph{IEEE Trans. Commun.}, vol.~74,
			pp.~2160--2175, 2025.
			\bibitem{tegos2025rate} S.~A. Tegos \emph{et al.}, ``Rate maximization for
			downlink pinching-antenna systems,'' \emph{IEEE Wireless Commun. Lett.},
			vol.~14, no.~5, pp.~1431--1435, 2025.
			\bibitem{xie2025lowcomplexity} X.~Xie, F.~Fang, Z.~Ding, and X.~Wang, ``A
			low-complexity placement design of PASS,'' \emph{IEEE Commun. Lett.},
			vol.~29, no.~8, pp.~1784--1788, 2025.
			\bibitem{zhao2026wdma} J.~Zhao \emph{et al.}, ``Waveguide division multiple
			access for PASS,'' \emph{IEEE Trans. Wireless Commun.}, vol.~25,
			pp.~13761--13775, 2026.
			\bibitem{uplinkact2026} Z.~Cheng and C.~Ouyang, ``Optimal uplink
			pinching-antenna activation,'' \emph{arXiv:2607.04159}, 2026.
			\bibitem{dualfed2026} X.~Xie, H.~Qin, F.~Fang, and X.~Wang, ``On dual-fed
			pinching antenna systems with in-waveguide attenuation,'' in
			\emph{Proc. IEEE GLOBECOM}, 2025.
			\bibitem{dingpoor2025blockage} Z.~Ding and H.~V. Poor, ``LoS blockage in
			pinching-antenna systems: Curse or blessing?'' \emph{IEEE Wireless Commun.
				Lett.}, vol.~14, no.~9, pp.~2798--2802, Sept. 2025.
			\bibitem{drl2026mobility} A.~Amhaz, M.~Elhattab, C.~Assi, and
			S.~Sharafeddine, ``Joint beamforming and antenna placement optimization in
			pinching antenna systems with user mobility: A deep reinforcement learning
			approach,'' \emph{arXiv:2605.08039}, 2026.
			\bibitem{sac2026umi} Q.~Zhao \emph{et al.}, ``Dynamic antenna placement for mobile users in urban micro
			pinching-antenna systems,'' \emph{arXiv:2603.00619}, 2026.
			\bibitem{constrained2026} M.~Jin \emph{et al.}, ``Constrained pinching antenna array design for sum-rate
			maximization in multi-user PASS,'' \emph{arXiv:2606.03830}, 2026.
\bibitem{dsp2026trajectory}
Y.~T.~Bozkurt, ``Trajectory-aware antenna reconfiguration for pinching antenna systems: Performance analysis under user mobility,'' \emph{Digital Signal Processing}, vol.~178, p.~106162, 2026.
			\bibitem{viterbi2025} V.~E. Galanopoulou \emph{et al.}, ``Viterbi state
			selection for discrete pinching antenna systems,''
			\emph{arXiv:2512.20389}, 2025.
			\bibitem{wu2020towards} Q.~Wu and R.~Zhang, ``Towards smart and
			reconfigurable environment: Intelligent reflecting surface aided wireless
			network,'' \emph{IEEE Commun. Mag.}, vol.~58, no.~1, pp.~106--112, 2020.
			\bibitem{li2024movementdelay} Q.~Li, W.~Mei, B.~Ning, and R.~Zhang,
			``Trajectory optimization for minimizing movement delay in movable antenna
			systems,'' \emph{IEEE Trans. Wireless Commun.}, 2025,
			doi:~10.1109/TWC.2025.3627982.
			\bibitem{lin2021perturbation} Y.~Lin \emph{et al.}, ``Perturbation-based
			regret analysis of predictive control in linear time-varying systems,'' in
			\emph{Proc. NeurIPS}, vol.~34, 2021, pp.~5174--5185.
	\end{thebibliography}}
\end{document}